\documentclass[12pt]{iopart}

\usepackage{iopams}
\usepackage{bm}
\usepackage{graphicx}
\usepackage{hyperref}
\begin{document}

\title[]{Machine Learning Inference of Molecular Dipole Moment in Liquid Water}

\author{Lisanne Knijff and Chao Zhang}

\address{Department of Chemistry-\AA ngstr\"om Laboratory, Uppsala
  University, L\"agerhyddsv\"agen 1, BOX 538, 75121, Uppsala, Sweden}
\ead{chao.zhang@kemi.uu.se}
\vspace{10pt}

\begin{abstract}
Molecular dipole moment in liquid water is an intriguing property, partly due to the fact that there is no unique way to partition the total electron density into individual molecular contributions. The prevailing method to circumvent this problem is to use maximally localized Wannier functions, which perform a unitary transformation of the occupied molecular orbitals by minimizing the spread function of Boys. Here we revisit this problem using a data-driven approach satisfying two physical constraints, namely:  i) The displacement of the atomic charges is proportional to the Berry phase polarization; ii) Each water molecule has a formal charge of zero. It turns out that the distribution of molecular dipole moments in liquid water inferred from latent variables is surprisingly similar to that obtained from maximally localized Wannier functions. Apart from putting a maximum-likelihood footnote to the established method, this work highlights the capability of graph convolution based charge models and the importance of physical constraints on improving the model interpretability.
\end{abstract}

%
%
%
%
%

Molecular dipole moment in polar liquids, such as liquid water, is an intriguing property ~\cite{debye1929polar}. On one hand, its magnitude and variation are directly linked to the dielectric properties of polar liquids and the intensities in infrared and sum-frequency generation spectra;  On the other hand, molecular dipole moment in polar liquids is not accessible to direct measurement.

One prominent example of using molecular dipole moment is the Kirkwood theory of the static dielectric constant $\epsilon $ in polar liquids~\cite{Kirkwood:1939br}, which directly involves the mean value of the molecular dipole in the liquid, i.e.~$\langle\mu\rangle$, and the orientational correlation factor, i.e.~the Kirkwood g-factor $g_{\rm{K}}$~\cite{Zhang:2016ho}.
\begin{equation}
\label{kirkwood_formula}
\frac{(\epsilon -1)(2\epsilon+1)}{\epsilon} = 4\pi \beta \rho\langle\mu\rangle^2 g_{\rm{K}}
\end{equation}
Here $\rho$ and $\beta$ are the number density and the inverse temperature respectively. In his 1939 paper,~\cite{Kirkwood:1939br} Kirkwood noticed that scaling the gas phase dipole of a water molecule by a factor of 1.26 may lead to an exact agreement on $\epsilon $ with the experiments and attributed this to a strong polarization effect by neighboring molecules in the liquid phase. 

The most notable estimation of the molecular dipole of water was given by Coulson and Eisenberg with an induction model in 1966~\cite{Coulson:ju}. They found that the reaction field from neighboring molecules in ice I$_h$ increases the molecular dipole to about 2.6 D, compared to 1.84 D of an isolated H$_2$O molecule. This issue was revisited in 1999~\cite{Silvestrelli:1999uu}. Instead of partitioning the total electron density of liquid water to individual water molecules~\cite{SITE:1999eza},  Silvestrelli and Parrinello found that $\langle\mu\rangle$ is about 3.0 D by using density functional theory based molecular dynamics (DFTMD)~\cite{Car:1985ix} and maximally localized Wannier functions (MLWFs)~\cite{Marzari:1997wa}. Their result agrees with the reanalysis of the Coulson-Eisenberg model~\cite{Batista:1998fb} and implies a charge transfer of about 0.5e along each O-H bond~\cite{Badyal:2000im,Liu:2016hz}. 

In MLWFs,~\cite{Marzari:1997wa} the Boys localization~\cite{Foster:1960gk} was employed to maximize the distance between centroids of orbitals (Wannier centers). However, alternative localization schemes are possible.~\cite{Lowdin:1966tq}. In a recent work,~\cite{Zhu:2020gp} Edmiston-Ruedenberg localization~\cite{Edmiston:1963fx}, which maximizes the self-repulsion energy of orbitals, has been explored to investigate the molecular dipole moment in liquid water. Depending on the regularization parameter, the resulting $\langle\mu\rangle$ varies between 2.35 D and 2.63 D~\cite{Zhu:2020gp}.  

In this work, we tackle this issue with a physically constrained data-driven approach. Similar to the dipole moment of an isolated molecule, the change in polarization $\Delta \mathbf{P}$ in condensed phase systems is well defined and experimentally measurable~\cite{Resta:2007jh}. For liquid water, with the choice of the molecular gauge, $\mathbf{P}$ is the so-called itinerant polarization~\cite{Caillol:1994ho}. This is the target quantity that we used in the regression task for liquid water. In contrast, molecular dipole moments here are inferred from latent variables and not involved in the training of the model. Another physical constraint built into our regression model is the charge neutrality of each water molecule, which is formally required by the integer change of the polarization quantum in the modern theory of polarization~\cite{KingSmith:1993hp,RESTA:1994cr}. Taking these ingredients into account, we show that the distribution of molecular dipole moments inferred from our regression model using the graph convolutional neural network architecture PiNet is surprisingly similar to that obtained from MLWFs. Moreover, the trained model, with PiNet using only data at ambient conditions, is transferable to liquid water in a range of different densities. 

The loss function $\mathcal{L}$ that we used for predicting the itinerant polarization $\mathbf{P}$ is the squared error in terms of the $l^2$ norm:
\begin{equation}
\label{eq:dip_loss}
    \mathcal{L} = \sum_i^n ||\mathbf{R}_i\mathbf{Q}_i-\mathbf{P}_i\Omega||^2_2
\end{equation}
where $\mathbf{R}_i$ is a $3\times N$ matrix of the atomic coordinates of the configuration $i$ for a system containing $N$ atoms. $\mathbf{Q}_i$ represents a column vector of the atomic charge in the configuration $i$. $\Omega$ is the volume of the simulation box. 

In our ML model, neutrality is enforced for each water molecule. This is done by subtracting the net charge using the following expression:
\begin{equation}
\label{eq:mol_charge_neutral}
    \mathbf{Q}_i = \bigoplus_{j=1}^{N\rm{w}} \left(\mathbf{Q}_{ij} - \frac{\mathbf{J}^\top\mathbf{Q}_{ij} \mathbf{J}}{3}\right)
\end{equation}
where $N_\textrm{w}$ is the number of water molecules in the system, $\mathbf{J}$ is a column vector with each entry equal to one, and $\bigoplus$ is used as the symbol for the concatenation of vectors. 

\begin{figure}[!ht]
\centering
\includegraphics[width=0.6\columnwidth]{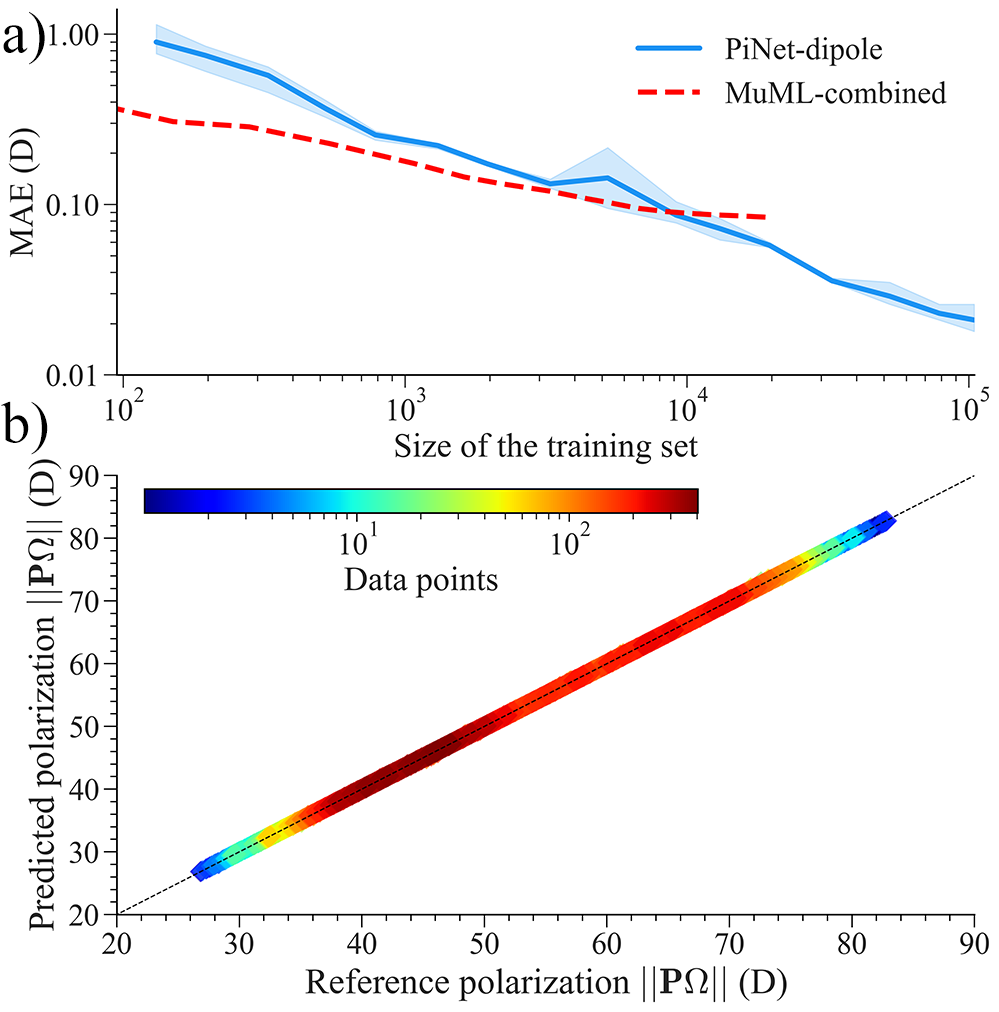}
     \caption{a) The learning curve of the PiNet-dipole model for the QM9 dataset in comparison with the MuML model which combines atomic charges and atomic dipoles. Data of the MuML model was extracted from Ref.~\cite{Veit:2020hw};  b) The parity plot for itinerant polarizations $\mathbf{P}$ of liquid water calculated using the Berry phase formula and the ones predicted from the PiNet-dipole model. }
    \label{fig:PiNet_prediction}
\end{figure}

The loss function in \Eref{eq:dip_loss} was trained with PiNet, a high-performing graph convolutional neural network architecture which works for both isolated molecules and condensed phase systems~\cite{Shao:2020kra}. Despite the simplicity of our scalar dipole model, PiNet leads to an outstanding learning curve with the QM9 dataset~\cite{ramakrishnan2014quantum}, compared to even more sophisticated models~\cite{Veit:2020hw} (see \Fref{fig:PiNet_prediction}a). For liquid water, the itinerant polarization $\mathbf{P}$ was computed with the CP2K package~\cite{Hutter:2013iea} and the BLYP functional~\cite{becke88,Lee:1988fm} using the Berry phase formula~\cite{RESTA:1998hj} (see Equation~(S1)). In this case, the PiNet-dipole model gives an excellent prediction of $||\mathbf{P}\Omega||$ with a mean absolute error (MAE) of 0.037 D for the test set (see \Fref{fig:PiNet_prediction}b). Here we followed the conventional 80:20 splitting of the dataset (see Electronic Supplementary Information for further descriptions of the computational methods and the dataset).  

Before looking into the inferred molecular dipole moment in liquid water from the PiNet-dipole model, it is necessary to set up a baseline for the sake of comparison. The baseline model used here is the linear regression (LR)-dipole model. 

\Eref{eq:dip_loss} resembles a LR problem where $\mathbf{Q}$ may be considered as the weight vector. However, this requires the configuration-dependent $\mathbf{Q}_i$ to be substituted by the configuration-independent $\bar{\mathbf{Q}}$. This leads to the loss function used in the LR-dipole model as
\begin{equation}
\label{eq:LS_Tikhonov_wat}
    \mathcal{L} = \sum_i^n ||\mathbf{R}_i \bar{\mathbf{Q}}  - \mathbf{P}_i\Omega||^2_2 + \lambda\sum_j^{N\textrm{w}} ||\mathbf{\Gamma}_j \bar{\mathbf{Q}} ||^2_2
\end{equation}
where Tikhonov matrices $\mathbf{\Gamma}$ are introduced to ensure the charge neutrality of each water molecule and $\lambda$ is the ridge parameter. For example, if the first three entries in $\bar{\mathbf{Q}}$ are  $\{q_\textrm{O}, q_\textrm{H}, q_\textrm{H}\}$ for one water molecule, then $\mathbf{\Gamma}_j$ can be written as:
\begin{equation}
\label{eq:Tikhonov_wat}
\mathbf{\Gamma}_j = 
\left [\begin{array}{c c c c c}
1 & 1 & 1 & 0 & \cdots \\
0 & 0 & 0 & 0 &\cdots  \\
\vdots & \vdots  & \vdots  & \vdots  & \vdots  
\end{array} \right]
\end{equation}

Therefore, the corresponding optimal solution for $\bar{\mathbf{Q}}^*$ is:
\begin{equation}
\label{eq:LS_solution_wat}
\bar{\mathbf{Q}}^* = \left(\mathbf{R}^{\top} \mathbf{R} + \lambda\sum_j^{N\textrm{w}} \mathbf{\Gamma}_j^\top \mathbf{\Gamma}_j\right)^{-1}\mathbf{R}^{\top}\mathbf{P}\Omega
\end{equation}
Note that here the subscript $i$ is dropped and that $\mathbf{R}$ includes all the input data over three Cartesian coordinates, which make it a $3n \times N$ matrix instead.

\begin{table}[!ht]
\centering
 \caption{The comparison between the linear regression (LR)-dipole model and the PiNet-dipole model for predicting the itinerant polarization $\mathbf{P}$ of liquid water in terms of the root mean square error (RMSE) and the mean absolute error (MAE). Entries are $||\mathbf{P}\Omega||$ in Debye (D). CN means applying the charge neutrality constraint for each water molecule.}
  \label{tab:LR_vs_PiNet}
  \begin{tabular}{ c c c}
    \hline
    Method & RMSE & MAE \\
    \hline
    LR-dipole (without CN)& 1.996 & 1.609
 \\ 
    LR-dipole& 2.033 & 1.632 \\
    PiNet-dipole (without CN) & 0.212 & 0.164
 \\ 
    PiNet-dipole & 0.054 & 0.037 \\
    \hline
  \end{tabular}
 
\end{table}

As seen in Table~\ref{tab:LR_vs_PiNet}, the itinerant polarization of liquid water predicted from the PiNet-dipole model is orders of magnitude more accurate than that predicted using the LR-dipole model. Moreover, the PiNet-dipole model with charge neutrality constraint performs significantly better than the one without charge neutrality constraint. These highlight the necessity of having environmental-dependent atomic charges built into the PiNet-dipole model and to impose physical constraints which are compatible with the modern theory of polarization.

\begin{figure}[!ht]
    \centering
    \includegraphics[width=0.6\columnwidth]{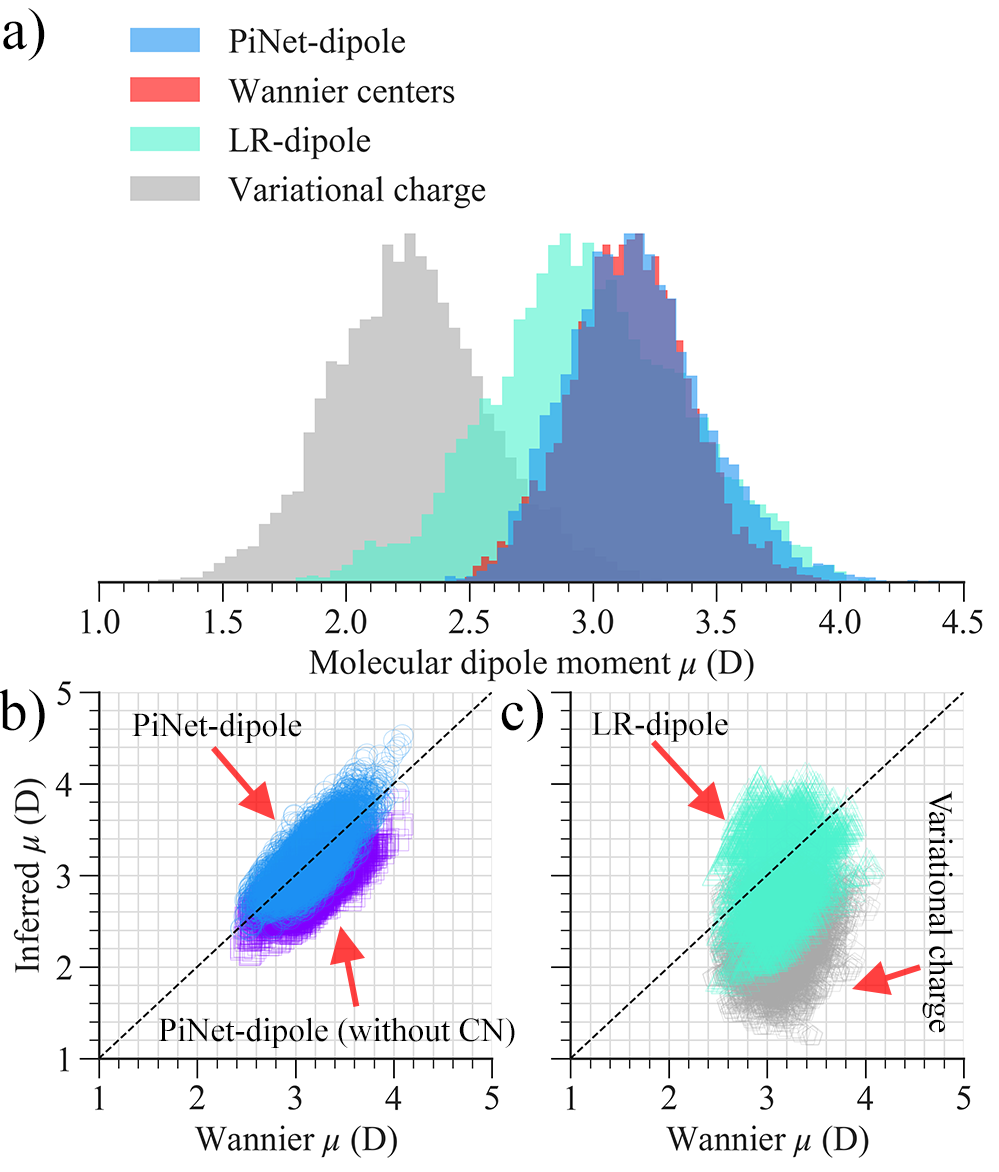}
    \caption{a) Distributions of molecular dipole moments in liquid water, as computed from Wannier centers and the variational charge method, and as inferred from the linear regression (LR)-dipole model and the PiNet-dipole model;  b) The parity plot for molecular dipole moments in liquid water between the ones computed from Wannier centers and the ones inferred from the PiNet-dipole models (with and without charge neutrality constraint); c) The parity plot for molecular dipole moments in liquid water between the ones computed from Wannier centers, and the ones inferred from the LR-dipole model and the variational charge method.}
    \label{fig:moldip_comparison}
\end{figure}

Now we are ready to compare the molecular dipole moments in liquid water as inferred from the PiNet-dipole model with those inferred from the linear regression dipole model. To put this comparison into perspective, we included the ones computed with Wannier centers from MLWFs (see Equation~(S2)). The reference calculations of the Wannier dipole moments were done with the same computational setup as the one used in computing the itinerant polarization $\mathbf{P}$ (see Electronic Supplementary Information for details). Results of this comparison are given in \Fref{fig:moldip_comparison}.

The most interesting finding in \Fref{fig:moldip_comparison}a is that the distribution of molecular dipole moments in liquid water inferred from the PiNet-dipole model is very close to that computed with Wannier centers. Since there is no regularization term in \Eref{eq:dip_loss}, this result may be seen as a maximum-likelihood footnote to MLWFs. In this regard, the LR-dipole model is quite off. In addition, as shown in \Fref{fig:moldip_comparison}b, the charge neutrality constraint is an important factor to achieve a good correlation. Because reproducing the itinerant polarization of liquid water and retaining the charge neutrality for each water molecule are two main ingredients in the PiNet-dipole model, the question that naturally arises is whether methods which satisfy these two conditions will necessarily lead to molecular dipole moments which are close to the ones obtained with Wannier centers in MLWFs.

Instead of viewing \Eref{eq:dip_loss} as a regression problem, one may use reference charges $\mathbf{Q}^\circ$ to supplement the equation and transform it to a set of linear equations~\cite{Laio:2004fx,Kirchner:2004fa}. This approach is what we call the variational charge method here, as the loss function shown below may be viewed as an energy functional with respect to the atomic charge $\mathbf{Q}_i$. 
\begin{equation}
\label{eq:varcharge_neutral}
    \mathcal{L} = \bm{\epsilon}^\top\left(\mathbf{R}_i\mathbf{Q}_i - \mathbf{P}_i\Omega\right) + \frac{\kappa}{2} || \mathbf{Q}_i-\mathbf{Q}^\circ_i||^2_2 
     + \bm{\eta}^\top\left(\sum_j^{N\textrm{w}}\mathbf{\Gamma}_j\mathbf{Q}_i\right)
\end{equation}
where $\bm{\epsilon}$ is a column vector which plays the role of a Lagrange multiplier and $\kappa$ is a weight parameter. To introduce charge neutrality for each water molecule, we added a third term in \Eref{eq:varcharge_neutral} where $\bm{\eta}$ is a Lagrange multiplier and $\mathbf{\Gamma}_j$ is the same Tikhonov matrix as given in \Eref{eq:Tikhonov_wat}.

Taking the derivative of \Eref{eq:varcharge_neutral} with respect to $\mathbf{Q}_i$, $\bm{\epsilon}$ and $\bm{\eta}$ leads to a set of $N+3+N/3$ linear equations:
\begin{equation}
\label{eq:varcharge_neutral_lin}
\left [\begin{array}{ c c c}
\kappa\mathbf{I} & \mathbf{R}_i^{\top} & \sum_j^{N\textrm{w}}\mathbf{\Gamma}_j^\top\\
\mathbf{R}_i& \mathbf{0} & \mathbf{0}  \\
\sum_j^{N\textrm{w}}\mathbf{\Gamma}_j & \mathbf{0} & \mathbf{0}
\end{array} \right]
\left[\begin{array}{c}
\mathbf{Q}_i\\
\bm{\epsilon}\\
\bm{\eta}
\end{array}\right]
= \left[ \begin{array}{c}
\kappa \mathbf{Q}^\circ\\
\mathbf{P}_i\Omega\\
\mathbf{0}
\end{array}\right]
\end{equation}
where $\mathbf{I}$ is a $N\times N$ identity matrix. Note that \Eref{eq:varcharge_neutral_lin} is solved independently for each configuration $i$.

The reference charges $\mathbf{Q}^\circ$ for liquid water were obtained with the REPEAT method \cite{Campana:2009bu,Golze:2015hf}, which provides electrostatic potential derived charges in periodic systems (see Electronic Supplementary Information for details). Since both $\bm{\epsilon}$ and $\bm{\eta}$ in \Eref{eq:varcharge_neutral} are Lagrange multipliers, the two physical constraints mentioned before (reproducing the itinerant polarization and retaining the charge neutrality) are exactly satisfied in the variational charge method. However, as shown in \Fref{fig:moldip_comparison}a and \Fref{fig:moldip_comparison}c, the resulting molecular dipole moments are very different from the ones obtained with Wannier centers. This suggests that one should not take the striking agreement between the molecular dipole moments in liquid water inferred from the PiNet-dipole model and those computed with Wannier centers for granted. The effective inclusion of long-range charge transfer in graph convolution~\cite{Zubatyuk:2019gp} is likely to be the reason behind this, because the delocalization tail along the hydrogen bonds contributes significantly to the molecular dipole moment in water clusters~\cite{Bako:2016ic}. 

\begin{figure}[!ht]
    \centering
    \includegraphics[width=0.6\columnwidth]{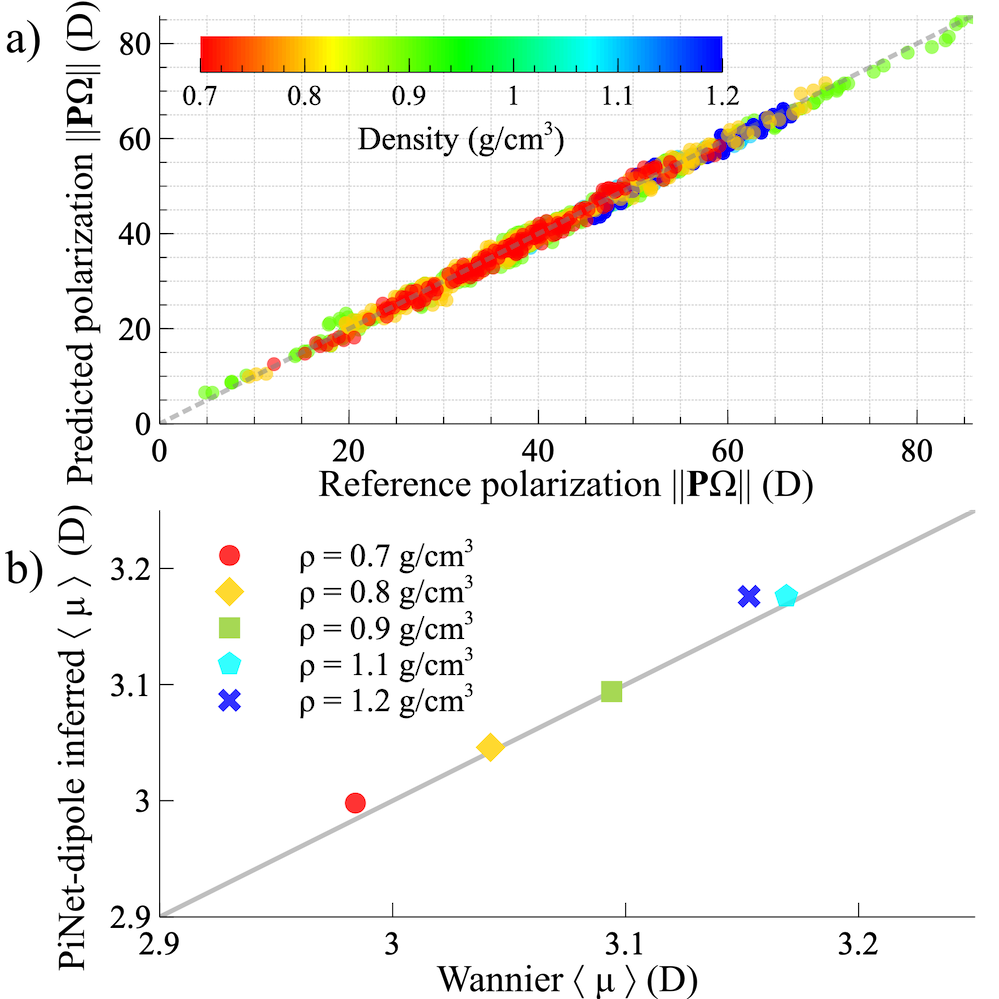}
    \caption{a) The parity plot for itinerant polarizations $\mathbf{P}$ of liquid water between ones computed from the Berry phase formula and ones predicted from the PiNet-dipole model at different densities; b) The parity plot for mean molecular dipole moments $\langle\mu\rangle$ in liquid water between the ones computed from the Wannier centers and the ones inferred from the PiNet-dipole model at different densities.}
    \label{fig:moldip_transfer}
\end{figure}

Finally, how good is the transferability of the PiNet-dipole model for liquid water? To answer that, we used a publicly accessible dataset for BLYP liquid water at different densities~\cite{Morawietz2019,Morawietz8368}. Results of the predicted itinerant polarizations $\mathbf{P}$ and the corresponding mean molecular dipole moments $\langle\mu\rangle$ are shown in \Fref{fig:moldip_transfer}. Despite the fact that the PiNet-dipole model was trained using data only at ambient conditions (i.e. $\rho=1.0$ g/cm$^3$), this model is rather transferable for liquid water in a range of different densities (\Fref{fig:moldip_transfer}a). Moreover, the mean water dipole moments $\langle \mu\rangle$ at different densities as inferred from the PiNet-dipole model correlate quite well with those calculated from the Wannier centers (figure~\ref{fig:moldip_transfer}b). These excellent agreements may be due to the high expressiveness of the underlying graph convolutional neural network architecture PiNet~\cite{Shao:2020kra}, which is also seen in the learning curve shown in \Fref{fig:PiNet_prediction}a for isolated molecules. 

Presumably, the paradigm in atomistic machine learning at present focuses on predicting physical quantities which enter into the loss function and the model training, e.g. energy, force and charge~\cite{Bartok:2017hz,AspuruGuzik:2018df,Lilienfeld:2020hz}. Instead, we looked into the latent variables of a trained network in this work and showed that the molecular dipole moments in liquid water inferred from the PiNet-dipole model are surprisingly in accord with those obtained from Wannier centers. Apart from the importance of physical constraints as shown in work, future studies should investigate other factors which could contribute to this agreement, e.g. an effective inclusion of long-range charge transfer. With these efforts, graph convolution based charge models could provide an alternative for describing the itinerant polarization in condensed phase systems~\cite{Zhang2020DeepInsulators}. Thus, their applications to charged systems, such as electrolyte materials and electrified solid-liquid interfaces~\cite{Zhang:2020ks,Shao:2021bf}, shall be anticipated. 

\section*{Data availability statement}

The polarization dataset for liquid water can be accessed using the following link \href{https://doi.org/10.5281/zenodo.4752246}{https://doi.org/10.5281/zenodo.4752246}.

\section*{Electronic Supplementary Information }
The Electronic Supplementary Information (ESI) includes: Description of QM9 and liquid water datasets, description of graph convolution and PiNet architecture, description of hyper-parameters used in the PiNet-dipole model, the comparison of charge distributions in liquid water from different models, and the influence of cutoff radius on the PiNet-dipole model.
\ack
 This project has received funding from the European Research Council (ERC) under the European Union's Horizon 2020 research and innovation programme (grant agreement No. 949012). L.K. is partly supported by a PhD studentship from the Centre for Interdisciplinary Mathematics (CIM) at Uppsala University. The calculations were performed on the resources provided by the Swedish National Infrastructure for Computing (SNIC) at UPPMAX and PDC.


\section*{References}
\providecommand{\newblock}{}

\end{document}


\maketitle

\makeatletter 
\renewcommand{\thefigure}{S\@arabic\c@figure} \renewcommand{\thetable}{S\@arabic\c@table}
\renewcommand{\theequation}{S\@arabic\c@equation} 
\renewcommand{\thepage}{S\arabic{page}}

\makeatother

\appendix{} 
\setcounter{figure}{0} 
\setcounter{table}{0}

\centerline{\bf\Large{Electronic Supplementary Information}}

\newpage 

\section{Description of the QM9 and the liquid water datasets}

\subsection{QM9}
The QM9 dataset is a subset of the GDB-17 database of 166 billion organic molecules \cite{ruddigkeit2012enumeration}.  The subset consists of 133885 molecules containing up to nine heavy atoms. These molecules contain carbon, oxygen, nitrogen, fluoride and hydrogen. Therefore, the dataset presents a general sample of the chemical space for small organic molecules. 3054 molecules were excluded from the dataset used in this work, as these failed the geometry consistency check. More detailed information about this dataset can be found in the original paper by Ramakrishnan et al.\cite{ramakrishnan2014quantum}.

\subsection{Liquid Water}
The liquid water dataset was created using a trajectory of \textit{ab initio} MD simulations of liquid water. Simulations were done using the CP2K package \cite{hutter2014cp2k,vandevondele2005quickstep}. A cubic box with a length of 12.432 \AA\ was filled with 64 water molecules. Then, an NVT simulation was performed at 330 K for 39000 steps with a timestep 0.5 fs, meaning a trajectory of 19.5 ps was generated. The BLYP functional \cite{becke1988density,lee1988development} was used with the TZV2P basis set \cite{vandevondele2007gaussian}, and the GTH pseudopotential \cite{goedecker1996separable}. The plane-wave cutoff was chosen as 600 Ry. The Bussi-Donadio-Parrinello thermostat \cite{bussi2007canonical} was used with a time constant of 20 fs. For every other frame in the trajectory the REPEAT charges \cite{campana2009electrostatic}, the Wannier centers~\cite{Marzari:1997wa}, and the Berry phase polarization \cite{KingSmith:1993hp,RESTA:1998hj} were computed using the CP2K package \cite{hutter2014cp2k,vandevondele2005quickstep}. Note that the itinerant polarization and molecular dipole moments can be readily obtained from Equation~\ref{eq:dipole_Wannier} by using the molecular gauge. Equivalently, the Berry phase formula (Equation~\ref{eq:berry_phase_RESTA}) also gives the itinerant polarization when the branch shift is taken into account. 

\begin{equation}
    \label{eq:berry_phase_RESTA}
    \mathbf{P}_\textrm{Berry}= \frac{1}{\Omega} \sum^{N_{ions}}_{i=1}Z_{i}\mathbf{R}_{i}-\frac{e}{2\pi L^2}\mathrm{Im}~ \mathrm{ln}\langle \Psi | e^{i(2\pi/L) \sum_i \hat{\mathbf{r}}_i} | \Psi \rangle
\end{equation}
where $Z_i$ and $\mathbf{R}_i$ are nuclear charge and position respectively. $\Psi$ is the many-body wave function, $e^{i(2\pi/L) \sum_i \hat{\mathbf{r}}_i}$ is the multiplicative operator and $L$ is the cell length of a cubic lattice (for the sake of simplicity) with the volume $\Omega$.



\begin{equation}
    \label{eq:dipole_Wannier}
    \mathbf{P}_\textrm{Wannier} = \frac{1}{\Omega}\left(\sum^{N_{ions}}_{i=1}Z_{i}\mathbf{R}_{i} -\sum^{N_{occ}}_{n=1}2e\Bar{\mathbf{r}}_{n}^{WC}\right)
\end{equation}

where $N_{ions}$ and $N_{occ}$ illustrates summing over the ions and the occupied states. Here, $\Bar{\mathbf{r}}_{n}^{WC}$ is the position of the Wannier center \cite{spaldin2012beginner}.

\section{Description of graph convolution and the PiNet architecture} \label{sec:PiNet}


A graph convolutional neural network (GCNN) combines the structure of a graph neural network and that of a convolutional neural network. A graph neural network ensures that the molecular structure is represented as a graph, where each atom is a node, and the pairwise interactions between the atoms are weighted edges. At each iteration, a message is passed between the nodes, which results in the node and edge feature vectors being iteratively updated. The convolutional part comes into play as a learnable radial filter that learns the features of the atoms within the cutoff radius $R_{c}$. Because of this, a hierarchy of features is created. The graph convolutional neural network has the same subnet for each element, as the generated node features are element specific \cite{shao2020pinn}. 

In this work, the PiNN library (\href{https://github.com/Teoroo-CMC/PiNN/}{https://github.com/Teoroo-CMC/PiNN/}) was used which was developed previously \cite{shao2020pinn}. PiNN is a Python library specifically designed to develop interpretable ANN architectures, and to implement existing ANNs in an efficient manner. To do this, PiNN describes neural networks through two different operations: the pairwise interaction (PI) operation, and the interaction pooling (IP) operation. PiNet uses these two operations to create an interpretable GCNN that describes pairwise interactions as a function of the distance between atoms. The form of the function is determined by the atomic properties of the interacting atoms. 

In the PI operation, the atomic properties $\textbf{P}_{i}$ can be used to create a pairwise interaction $\textbf{I}_{ij}$, which is a function of the starting properties of the two atoms as well as the distance $r_{ij}$ between them (see Equation \ref{eq:PI}). 

\begin{equation}
   \textbf{I}_{ij}^{t} = \textup{PI}(\textbf{P}_{i}^{t}, \textbf{P}_{j}^{t}, r_{ij})
   \label{eq:PI}
\end{equation}

In Equation \ref{eq:PI}, \textit{t} is the iterator. This iterator goes up to the number of graph convolutional blocks in the graph convolutional neural network. The use of the iterator ensures that many-body interactions are taken into account. 

The opposite of this is the IP operation. This operation can be used to compute a new atomic property by summing over all the pairwise interactions of a certain atom and passing this to a function, Equation \ref{eq:IP}. The result of this is an updated atomic property. Since all the interactions are summed over, this also ensures that the result is permutation invariant.

\begin{equation}
    \textbf{P}_{i}^{t+1} = \textup{IP}\left ( \sum _{j} \textbf{I}_{ij}^{t} \right )
    \label{eq:IP}
\end{equation}

Combining these two operators creates an atomic fingerprint that is capable of capturing the atomic information of the neighbors and using this to update the information of the atom. When multiple graph convolutional blocks are used $\textbf{I}_{ij}$ gets updated as well. 
In particular, PiNet expresses the PI operation as three parts. First, the interatomic distances are expressed in a radial basis $\textbf{e}_{ij}$. To calculate this, $n_\textup{basis}$ Gaussian functions multiplied with a cutoff function $f_{c}$ are used. As a cutoff function, the cutoff function by Behler et al. \cite{behler_atom-centered_2011} is used. 

Then, the activation occurs by applying the PI layer, which is a feed-forward neural network. This neural network generates a weight matrix $\textbf{W}_{ij}$ from the atomic properties $\textbf{P}_{i}$, and $\textbf{P}_{j}$. 

\begin{equation}
   \textbf{W}_{ij} = \textup{NN}^{\textup{PI-layer}}([\textbf{P}_{i}, \textbf{P}_{j}])
    \label{eq:wij}
\end{equation}

Finally, activation through the interaction-interaction (II) layer occurs. This II layer is another feed-forward neural network which uses the information from the previous two parts as input to compute the interaction $\textbf{I}_{ij}$.

\begin{equation}
    \textbf{I}_{ij} = \textup{NN}^{\textup{II-layer}}(\textbf{W}_{ij}\textbf{e}_{ij})
    \label{eq:Iij}
\end{equation}


In order to further refine the updated atomic property, another feed-forward neural network is used. This is done in the property-to-property (PP) layer. Afterwards, the information is passed to the next GC block \cite{shao2020pinn}.

\section{Description of the hyper-parameters used in the PiNet-dipole model}


The PiNet-dipole model used here has the following architecture: The II layer consists of four layers of 64 nodes. The PI layer consists of a single layer of 64 nodes. The PP layer consists of four layers of 64 nodes. The en-layer, which can be considered as an IP layer but with fully connected nodes and coefficients, consists of 64 fully connected nodes before the final result is produced. Five graph convolutional blocks were used in the network. Ten Gaussian basis functions were used which have a width of 3.0 \AA, and a cutoff radius $R_{c}$ of 4.5 \AA.  The charge neutral model was trained for 1.5 million steps, the model without charge neutrality constraint was trained for 2 million steps, and a batch size of 200 data points was used during the training. The learning rate was set to be $10^{-4}$ at the beginning and was scaled by 0.994 every 10000 steps.

\section{The comparison of charge distributions in liquid water from different models}

\begin{figure}[!ht]
    \centering
    \includegraphics[width=0.8\linewidth]{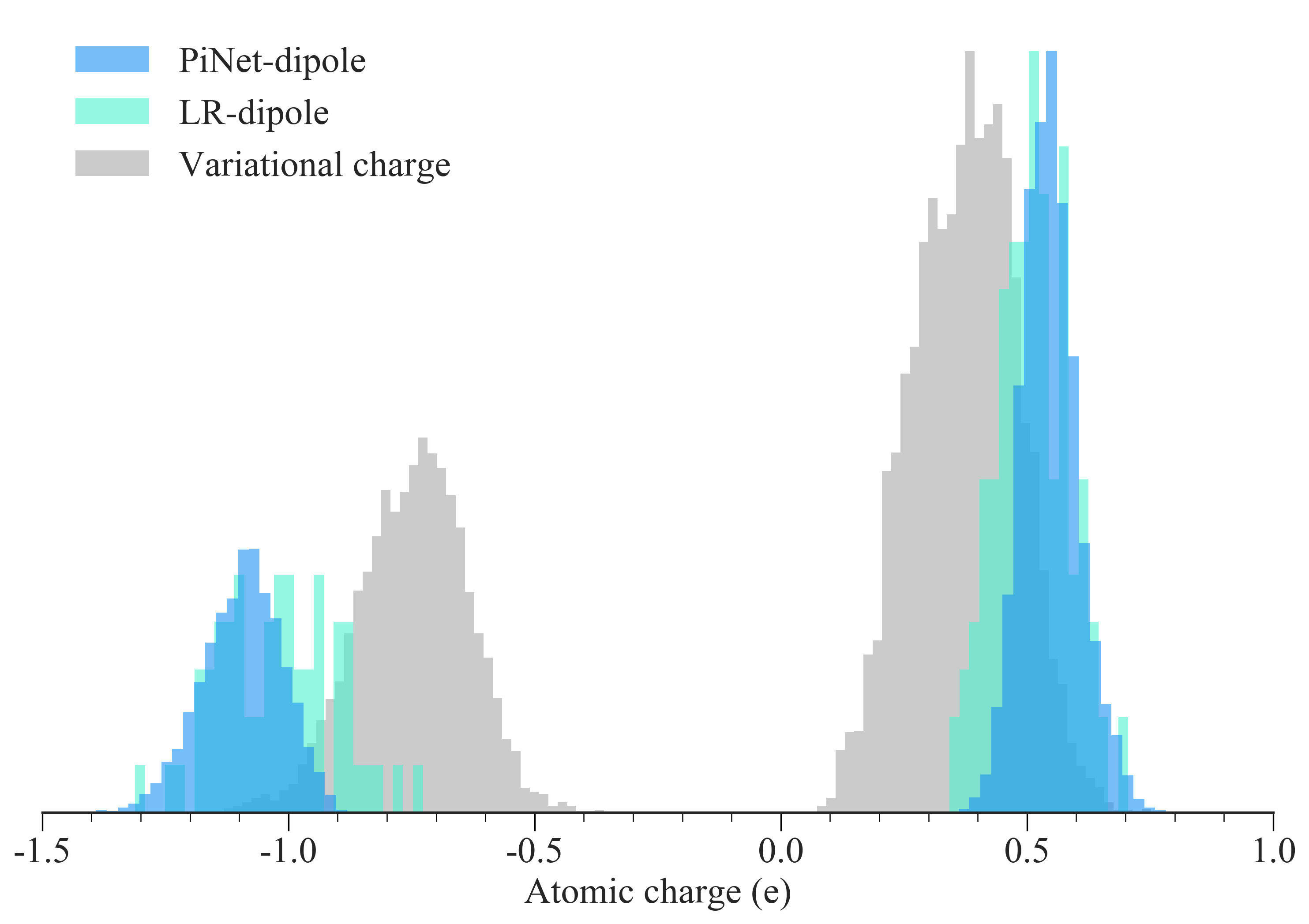}
    \caption{a) The distribution of charges of liquid water  predicted from the PiNet-dipole model, the LR-dipole model, and the variational charge method.}
    \label{fig:SI1}
\end{figure}

\section{The influence of the cutoff radius on the PiNet-dipole model}

\begin{figure}[H]
    \centering
    \includegraphics[width=\linewidth]{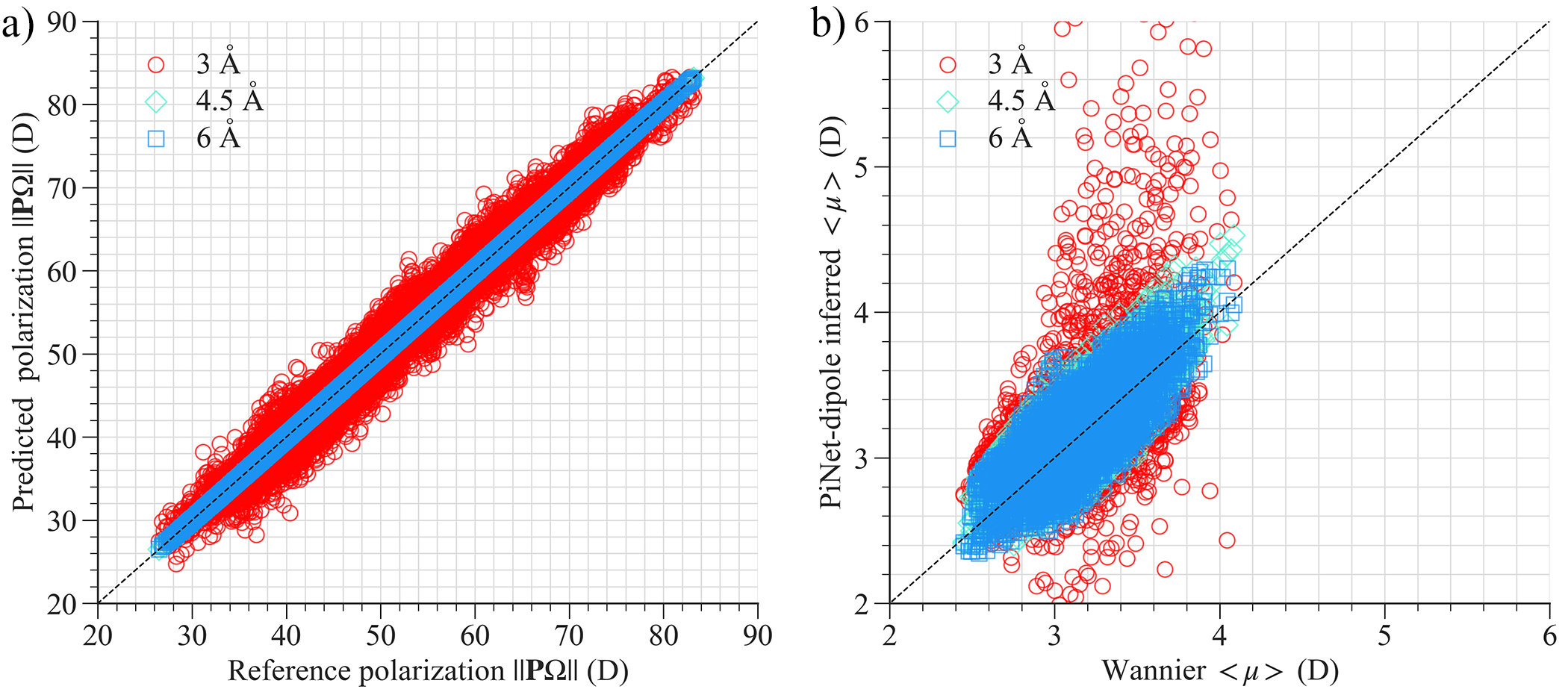}
    \caption{a) The parity plot for itinerant polarizations $\mathbf{P}$ of liquid water predicted from the PiNet-dipole model using a cutoff range of 3 \AA, 4.5 \AA, and 6 \AA. ; b) The parity plot for molecular dipole moments in liquid water predicted from the PiNet-dipole model using a cutoff range of 3 \AA, 4.5 \AA, and 6 \AA.}
    \label{fig:SI2}
\end{figure}